\documentclass[10pt]{iopart}

\usepackage{iopams,bm,array,epsfig,fullpage,stmaryrd}

\newenvironment{stabular}[1]
  {\tabular{#1}}
  {\endtabular}

\newcommand{\re}{\textrm{Re}}

\newcommand{\aS}{\llbracket S \rrbracket}
\newcommand{\aI}{\llbracket I \rrbracket}
\newcommand{\aR}{\llbracket R \rrbracket}
\newcommand{\aSS}{\llbracket SS \rrbracket}
\newcommand{\aSI}{\llbracket SI \rrbracket}
\newcommand{\aSR}{\llbracket SR \rrbracket}
\newcommand{\aII}{\llbracket II \rrbracket}
\newcommand{\aIR}{\llbracket IR \rrbracket}
\newcommand{\aRR}{\llbracket RR \rrbracket}

\newcommand{\be}{\begin{equation}}
\newcommand{\ee}{\end{equation}}
\newcommand{\bd}{\begin{displaymath}}
\newcommand{\ed}{\end{displaymath}}
\newcommand{\BE}{\begin{eqnarray}}
\newcommand{\EE}{\end{eqnarray}}

\begin{document}
\title[Stochastic oscillations of adaptive networks]{Stochastic oscillations of adaptive networks:\\ application to epidemic modelling}

\author{Tim Rogers, William Clifford-Brown, Catherine Mills, Tobias Galla}
\address{Theoretical Physics, School of Physics and Astronomy, The University of Manchester, Manchester M13 9PL, United Kingdom}
\begin{abstract}
Adaptive-network models are typically studied using deterministic differential equations which approximately describe their dynamics. In simulations, however, the discrete nature of the network gives rise to intrinsic noise which can radically alter the system's behaviour. In this article we develop a method to predict the effects of stochasticity in adaptive networks by making use of a pair-based proxy model. The technique is developed in the context of an epidemiological model of a disease spreading over an adaptive network of infectious contact. Our analysis reveals that in this model the structure of the network exhibits stochastic oscillations in response to fluctuations in the disease dynamic. 
\end{abstract}
\hspace{72pt}\begin{footnotesize}Email for correspondence: \mailto{tim.rogers@manchester.ac.uk}\end{footnotesize}

\section{Introduction}

The importance of network structure to the dynamics of interacting agents in the real world is now almost universally recognised. Traditionally, systems of interacting agents were modelled by means of simple ordinary differential equations, describing the deterministic `mean-field' behaviour of well-mixed populations. Whilst this approach has been very successful in the past (examples can be found in ecology, evolution and game theory \cite{Proulx, Szabo}), it is valid only in simplified models in which each agent is equally likely to interact with any other agent. This caveat places a severe constraint on the applicability of these techniques to real-world systems.

In recent years it has become possible to go beyond models of well-mixed populations and to move towards the incorporation of network structure. This step forward has been facilitated by several recent advances: the increasing quality and availability of data on naturally occurring networks; the development of a more comprehensive theory of complex networks (see e.g. \cite{newman}); and lastly, the exponential growth in computing power allowing for large-scale simulations of network models. In these models, individual agents are placed on the nodes of a network\footnote{We are here only concerned with models in which there is one single individual at each node of the network. An example of so-called meta-population models, with multiple individuals at each node, can be found in \cite{Rozhnova2011}.} and interactions occur across the edges of the network. Broadly speaking there are two classes of individual-based models on networks (i) those in which the underlying network is assumed to be static and (ii) models which take into account changes to the structure of the network concurrently with the agent dynamics. The former class focusses on the dynamics \emph{on} the network and is a good proxy when the evolution of the network structure is slow compared to the dynamics of the nodes. The latter approach applies more generally as it considers dynamics both \emph{on} and \emph{of} the network, occurring simultaneously. 

This second approach is frequently referred to as the  `adaptive network' modelling paradigm, and it is an important recent development, providing a useful framework to model many real-world systems \cite{Gross2009}. The conceptual simplicity of these models is, however, countered by the fact that it is notoriously difficult to develop a succinct mathematical theory for their macroscopic behaviour. The equilibrium properties of adaptive-network models have been studied in depth \cite{Gross2009}, frequently by means of the so-called pair approximation (PA), usually attributed to \cite{Keeling1997}. This approach is widely used in adaptive networks; although the motivation and quality of the approximation are questionable \cite{Rogers2011}, results (for equilibria at least) are typically reasonable.

The mathematical analysis of the actual dynamics of adaptive-network models and their approach to equilibrium is intricate; in particular, the effects of what is referred to as `intrinsic noise' or `demographic stochasticity' are not at all well-understood.  Such stochasticity can have a profound impact on the dynamics of individual-based models; a realisation which carries with it important implications for fields such as ecology \cite{McKane2005,Rogers2012} and epidemiology \cite{Alonso2007}. A robust theoretical framework with which to describe effects of intrinsic noise has been developed for well-mixed models, primarily based around the system-size expansion method of van Kampen \cite{vanKampen1992}, but no comprehensive approach exists for models on adaptive networks. 

It is the purpose of this article to propose a general theoretical method to address effects of intrinsic noise in adaptive-network models. Starting from the rules of the adaptive-network model, we derive a low-dimensional Markov jump process which (approximately) captures the aggregate dynamics of the agents and edges in the network model. We refer to this model as the `pair-based proxy' (PBP). In deriving the PBP we employ a moment-closure assumption equivalent to the PA. Our implementation is somewhat different to usual in that we are not attempting to derive a set of ordinary differential equations; instead we define a new stochastic process designed to reproduce the macroscopic behaviour of the original networked model. In doing so we gain instant access to established methods applicable to jump processes. In the limit of infinitely large networks the jumps in the PBP become infinitesimally small, and the dynamical rules reduce to a system of differential equations. Stochastic effects in large but finite networks can be investigated using the linear-noise approximation \cite{vanKampen1992}. 

One particularly important application of adaptive-network models is to the study of infectious disease, and we have chosen to develop our method in this context. Epidemiological models involving well-mixed populations, or even static networks, are at best a crude approximation to the dynamics of real-world epidemics. Adaptive-network models represent a step towards greater realism, capturing the dynamics of a disease which spreads over a constantly evolving network of infectious contact between individuals. Moreover, adaptive networks offer the possibility to model the effects of intervention strategies aimed at disrupting the network of infection, such as contact tracing and quarantining. It is also becoming clear that demographic noise has an important role to play in the dynamics of epidemics, particularly the noise-induced excitation of transient oscillations, leading to cycles of disease outbreaks \cite{Kuske2007,Alonso2007}. 

The first adaptive network epidemic model is due to Gross \textit{et al.} \cite{Gross2006}, who studied the equilibrium properties of the model by means of the PA (they provide a description of the full phase diagram, including an active non-equilibrium phase, together with an analysis of the structural properties of the network). This deterministic approach is successful in describing stationary states and phase behaviour, but it systematically neglects effects of noise. This work was followed by Shaw and Schwartz \cite{Shaw2008} who added extrinsic noise `by hand' to the deterministic equations. The method of Rozhnova and Nunes \cite{Rozhnova2009} addresses intrinsic fluctuations due to demographic stochasticity, however, it is only applicable to static networks in which each node has the same number of neighbours. Our approach is similar to that of \cite{Rozhnova2009}, who reverse-engineered a stochastic pair-based model to fit the macroscopic PA equations obtained from the network model, but crucially we here address the case of adaptive networks. We expect that the general scheme we introduce will be useful in quantifying stochastic effects in a wide range of adaptive-network models, not only in epidemiology but in other areas as well. As a key result of our analysis we are able to identify oscillations of the underlying network structure itself, induced by the stochasticity of the dynamics. The (approximate) mapping onto the PBP allows us to make analytical predictions for the spectral properties of these oscillations of the network, as well as for those of the population of agents on the network.

The remainder of this paper is organised as follows: In the next section we define an adaptive-network model of an epidemic and show how to derive the low-dimensional PBP model which approximates its dynamics. The next section contains a theoretical analysis of the PBP for large networks, in which expressions for deterministic dynamics and stochastic corrections are computed. In section 4, we compare the theoretical results from the PBP with simulations of the adaptive-network model and discuss some of the limitations of the approximation. In section 5 we report on the observation that the structure of the network itself undergoes noise-driven oscillations, and show how this behaviour can be captured by the PBP. 

\section{Constructing the pair-based proxy}
\subsection{Adaptive network model}
We consider a susceptible-infected-recovered-susceptible (SIRS) disease model in a cohort of $N$ individuals, which are placed at the nodes of a network and joined by a total of $K$ edges. The edges of the network represent the potential for infectious contact over which the disease may spread: whenever a susceptible ($S$) individual shares an edge with an infected ($I$) individual, the susceptible individual becomes infected with rate $\beta$. Infected individuals become recovered ($R$) with rate $\gamma$, and recovered individuals become susceptible again with rate $\delta$. If we label individuals by $i=1,\dots,N$, and if $a_{ij}\in\{0,1\}$ is the adjacency matrix of the network (we consider only undirected networks, i.e. $a_{ij}=a_{ji}$), then the dynamics can be described by the reactions
\begin{eqnarray}
&S_i+I_j \stackrel{\beta a_{ij}}{\longrightarrow} I_i+I_j\,, \qquad I_i \stackrel{\gamma}{\longrightarrow} R_i \,,\qquad R_i \stackrel{\delta}{\longrightarrow} S_i,
\end{eqnarray}
where the notation $S_i$ indicates that the individual at node $i$ is in the susceptible state (and similar for $I_i$ and $R_i$).

In the real world, the transmission of a disease requires spatial proximity, and thus the network of potential infection is in a constant state of flux caused by the movement of individuals. We incorporate this fact into our model through a process of constant low-level re-arrangement. Each edge in the network decays (i.e. is removed) with rate $\mu$, at which point a new edge is generated between a random pair of individuals elsewhere in the network\footnote{The total number of edges in the network remains constant under this rewiring process. It is also possible to consider edge creation and deletion as separate processes so that the total number fluctuates, however, this choice does not make a qualitative difference to the aggregate behaviour of the model.}. \par
As well as this random rewiring process, the structure of the network adapts intelligently in response to the progression of the epidemic by a process of `smart' rewiring. For each susceptible--infected pair there is a chance that the susceptible individual will discover the infection and choose to remove that edge and replace it with an edge to a randomly chosen susceptible individual. This happens with rate $w$ for each $SI$ pair.\par
When referring to the state of the networked model, we write $[X]$ for the number of nodes in state $X$ and $[XY]$ for the total number of edges in the system between pairs of nodes in states $X$ and $Y$, for any $X,Y\in\{S,I,R\}$. Later we will discuss the behaviour of the model in the limit of large network size, by which we mean that $N\to\infty$ and $K\to\infty$ with $N/K$ held constant.
\subsection{Pair-based proxy model}
As discussed in the introduction, we intend to study the macroscopic behaviour of the adaptive-network model through a low-dimension model which captures the main features of the network model whilst remaining analytically tractable. We will define a 9-dimensional Markov jump process with state vector $$\bm{X}=(\aS\,\,\, \aI\,\,\, \aR\,\,\, \aSS\,\,\, \aSI\,\,\, \aSR\,\,\, \aRR\,\,\, \aIR\,\,\, \aII)^T\,.$$ 
The components of the state vector are intended to represent the corresponding quantities in the network model: our goal is to derive the transition probabilities for the jump process from the network model in such a way that the statistics of, for example, $\aS$ and $\aSI$ in the PBP model will reflect those of $[S]$ and $[SI]$ in the original network model. 

For each of the five processes taking place in the network model (i.e. infection, recovery, rebirth, random rewiring and smart rewiring) we will compute the average net change to each $[X]$ and $[XY]$, and use this to specify the jump direction and magnitude in the PBP. The jump rates are also drawn from the network model in this way. To illustrate this construction, we discuss in detail the process of infection. Our argument proceeds along the following steps:
\begin{enumerate}
\item[1.] The rate with which an infection event takes place in the network model is $\beta [SI]$. The rate for infection events in the PBP model is therefore set to $\beta \aSI$.
\item[2.] When an infection takes place in the network, one $S$ individual is converted into an $I$. For the PBP we write this as a jump with direction $S\mapsto I$ and magnitude 1, meaning that $\aS\mapsto\aS -1$ and $\aI\mapsto \aI +1$.
\item[3.] An infection event in the network model changes the configuration of the system on the level of edges as well, and several edges will be affected, namely those between the newly infected individual and its neighbours. So an infection event will generally result in one (or several) $SI$ edges becoming $II$ edges, and similarly some $SR$ edges may become $IR$, and some $SS$ may become $SI$.  It is not obvious at first sight how many of each of the different types of edges are changed during an infection event. 
\item[4.] For the purposes of the PBP model we use the {\em typical} change to $[XY]$ which takes place in the network model during an infection event. In order to compute these it is first necessary to determine the average number of neighbours an $S$-individual has in the network:
\be
[k_S]=\frac{2[SS]+[SI]+[SR]}{[S]}.
\ee
In the PBP model this quantity is approximated by $\llbracket k_S\rrbracket=(2\aSS+\aSI+\aSR)/\aS$.
\item[5.] Next we determine how many neighbours of a newly-infected individual we expect to be in states $S$, $I$ and $R$ respectively. In the network model, the probability that a randomly chosen neighbour of an $S$-individual is of type $S$ (or $I$ or $R$) is $2[SS]/([k_S][S])$ (or $[SI]/([k_S][S])$ or $[SR]/([k_S][S])$, respectively). 
\item[6.] In the network model the states of the individuals surrounding the individual of type $S$ are correlated. In order to formulate the PBP we assume that only correlations via the central node are relevant. This is equivalent to the usual pair moment closure assumption used in the literature \cite{Keeling1997,Gross2006,Shaw2008,Rozhnova2009}. Making this assumption, the magnitude of the jump in the direction $SS\mapsto SI$ is $2z\aSS/\aS$, where $z=(\llbracket k_S\rrbracket -1)/\llbracket k_S\rrbracket$. Similarly, the jumps in directions $SI\mapsto II$ and $SR\mapsto IR$ have magnitudes $1+z\aSI/\aS$ and $z\aSR/\aS$, respectively.
\item[7.] Each of the jumps computed above are taken to happen simultaneously; the combined jump is calculated by vector addition. To summarize, infection events in the PBP model occur with rate $\beta \aSI$ and lead to the changes to the state vector $\bm{X}$ as indicated in the first column of the stoichiometric matrix in Eq. (\ref{eq:matrix}) below.
\end{enumerate}
A full list of the transition events, rates and state changes is shown in Table \ref{PBP}. The column labelled `jump rate' indicates how many events of the different types (infection, recovery, rebirth, smart rewiring and random rewiring) occur per unit time in the PBP. For clarity, in the `jump direction' and `jump magnitude' columns we have split the jumps into individual contributions of the form $X\mapsto Y$ or $XZ\mapsto YZ$. Note that the jump magnitudes for the edges are typically not integer.
\begin{table}
\begin{center}
\begin{stabular}{m{40mm}m{45mm}m{33mm}m{30mm}}
\textbf{Event} & \textbf{Jump rate} & \textbf{Jump direction} & \textbf{Jump magnitude} \\\hline\hline
Infection & $\beta \aSI $ & $S\mapsto I$\par $SS\mapsto SI$ \par $SI\mapsto II$ \par $SR\mapsto IR$ & 1 \par $2z\aSS /\aS $\par $1+z\aSI /\aS $ \par $z\aSR /\aS $\\\hline
Recovery & $\gamma\aI $ & $I\mapsto R$\par $SI\mapsto SR$ \par $II\mapsto IR$ \par $IR\mapsto RR$ & 1 \par $\aSI /\aI $\par $2\aII /\aI $ \par $\aIR /\aI $\\\hline
Rebirth & $\delta\aR $ & $R\mapsto S$\par $SR\mapsto SS$ \par $IR\mapsto SI$ \par $RR\mapsto SR$ & 1 \par $\aSR /\aR $\par $\aIR /\aR $ \par $2\aRR /\aR $\\\hline
Smart rewiring& $w\aSI $ & $SI\mapsto SS$ & 1\\\hline
Random rewiring\par ($A,B,C,D\in\{S,I,R\}$)& $\mu (2-\delta_{C,D}) \llbracket AB\rrbracket \llbracket C\rrbracket \llbracket D\rrbracket/N^2 $ & $AB\mapsto CD$&1 \\\hline
\end{stabular}
\end{center}
\caption{\label{PBP}List of reactions in the pair-based microscopic model. Here $z=(\llbracket k_S\rrbracket-1)/\llbracket k_S\rrbracket$, where $ \llbracket k_S\rrbracket=(2\aSS +\aSI +\aSR )/\aS $ is the average degree of a susceptible node.}
\end{table}

The effects of infection, recovery, rebirth and smart rewiring are also summarized in the following stoichiometric\footnote{We are slightly abusing terminology here, as some entries of our `stoichiometric' matrix are not integers, and moreover they depend on the state of the system. This is not the normal convention.} matrix:
\begin{equation}\label{eq:matrix}
\fl\qquad\mathbb{T}^{(1)}=\left(\begin{array}{cccc} -1 & 0 & 1 & 0\\ 1 & -1 & 0 & 0 \\ 0 & 1 & -1 & 0 \\ -2z\aSS /\aS  & 0 & \aSR /\aR  & 1 \\ z\big(2\aSS -\aSI \big)/\aS -1 & -\aSI /\aI  & \aIR /\aR  & -1\\ -z\aSR /\aS  & \aSI /\aS  & \big(2\aRR -\aSR \big)/\aR  & 0\\ z\aSI /\aS +1 & -2\aII /\aI  & 0  & 0\\ z\aSR /\aS  & \big(2\aII -\aIR \big)/\aI  & -\aIR /\aR   & 0\\ 0 & \aIR /\aI  & -2\aRR /\aR  & 0\end{array}\right).
\end{equation}
The first column for example indicates that the number of individuals of type $S$ is reduced by one in the event of an infection, the number of $I$ increases by one, the number of $R$ remains unchanged, and the number of particles of type $SS$ is reduced by $2z\aSS/\aS$, and so on. The second, third and fourth columns of the above matrix describe recovery, rebirth and smart rewiring events; the corresponding entries can read off from the table. The rates with which infection, recovery, rebirth and smart rewiring occur are summarized in the rate vector
\begin{equation}\label{eq:rv1}
\bm{R}^{(1)}=\left(\begin{array}{c}\beta\aSI \\\gamma\aI \\\delta\aR \\w\aSI \end{array}\right)\,.
\end{equation}
The random rewiring process is composed of 36 separate reactions, and the associated stoichiometric  matrix $\mathbb{T}^{(2)}$ (of dimensions $9\times 36$) and rate vector $\bm{R}^{(2)}$ are thus rather large. They are constructed as follows: for each combination $A,B,C,D\in\{S,I,R\}$ the corresponding column of $\mathbb{T}^{(2)}$ has a $-1$ in position for $\llbracket AB\rrbracket $, a $+1$ in the position for $\llbracket CD\rrbracket $, and zeros elsewhere; the corresponding entry of the rate vector is $\mu(2-\delta_{C,D})\llbracket AB\rrbracket \llbracket C\rrbracket \llbracket D\rrbracket/N^2$. The stoichiometric matrix and rate vector for the whole model are found by concatenating $\mathbb{T}^{(1)}$ with $\mathbb{T}^{(2)}$ and $\bm{R}^{(1)}$ with $\bm{R}^{(2)}$. We write
\begin{equation}
\mathbb{T}=\left(\begin{array}{cc}\mathbb{T}^{(1)}&\mathbb{T}^{(2)}\end{array}\right)\,,\qquad \bm{R}=\left(\begin{array}{c}\bm{R}^{(1)}\\\bm{R}^{(2)}\end{array}\right),
\end{equation}
where $\mathbb{T}$ is now a matrix of dimensions $9\times 40$.

The initial condition for the PBP is taken to match that of the network model (i.e. $\aS=[S]$, $\aI=[I]$, and so on). Note that the sum of the first three entries in each column of $\mathbb{T}$ is zero, as is the sum of the remaining entries in each column. This fact implies the conditions $\aS+\aI+\aR=N$ and $\aSS+\aSI+\aSR+\aII+\aIR+\aRR=K$ (which are inherited from the initial condition of the network model) must hold throughout the dynamics. 

\section{Analysis of the pair-based model}
\subsection{Effective Langevin equation}
We now proceed to derive an analytical description of the PBP model in the limit of large $N$. The first step is to introduce the density vector  $\bm{x}=\bm{X}/N$ and rescaled rate vector $\bm{r}=\bm{R}/N$. There is no need to adjust the stoichiometric matrix though, as its entries are invariant under scaling of $\bm{X}$. The scaled variable $\bm{x}$ is a realisation of a density dependent Markov process satisfying the conditions of Kurtz' theorem \cite{Kurtz1978}. Employing that result, we can immediately write down a Langevin equation for $\bm{x}$ (to be interpreted in the It\={o} sense) which holds when $N$ is large:

\begin{equation}
\frac{d}{dt}\bm{x}=\mathbb{T}\,\bm{r} + \frac{1}{\sqrt{N}}\bm{\eta}(t)\,,
\label{langevin}
\end{equation}
where $\bm{\eta}$ is a vector of multiplicative Gaussian noise variables with state-dependent correlator 
\begin{equation}
\big\langle \eta_i(t)\eta_j(t') \big\rangle=\delta(t-t')\sum_{k=1}^{40}\mathbb{T}_{ik}\bm{r}_k\mathbb{T}_{kj}.
\end{equation}
The variable $k$ here runs through all $40$ reactions of the PBP model. This formulation is equivalent to that obtained by the performing a Kramers-Moyal expansion \cite{vanKampen1992} of the Master equation and neglecting terms including and beyond $1/N$.
\subsection{Deterministic limit}
In the limit of large population size (that is, $N\gg1$, $K\gg1$ with constant $K/N$) Eq. (\ref{langevin}) reduces to a deterministic system of differential equations given by $\dot{\bm{x}}=\mathbb{T}\bm{r}$. Returning to the original scaling, the deterministic equations are:
\begin{eqnarray}\label{deteqns}
\fl\qquad\frac{d\aS}{dt}=\delta\aR-\beta\aSI,\nonumber\\
\fl\qquad\frac{d\aI}{dt}=\beta\aSI-\gamma\aI,\nonumber\\
\fl\qquad\frac{d\aR}{dt}=\gamma\aI-\delta\aR,\nonumber\\
\fl\qquad\frac{d\aSS}{dt}=\delta\aSR  -2z\beta\frac{\aSI\aSS}{\aS}+w\aSI-\mu\aSS + K\mu\frac{\aS^2}{N^2},\nonumber\\
\fl\qquad\frac{d\aSI}{dt}=\beta\aSI \left(2z\frac{\aSS}{\aS}-z\frac{\aSI}{\aS}-1\right)+\delta\aIR-\aSI\gamma-(w+\mu)\aSI + 2K\mu\frac{\aS\aI}{N^2},\nonumber\\
\fl\qquad\frac{d\aSR}{dt}=\gamma\aSI+\delta\left(2\aRR-\aSR\right)-z\beta\frac{\aSI\aSR}{\aS}-\mu\aSR +2K\mu\frac{\aS\aR}{N^2},\nonumber\\
\fl\qquad\frac{d\aII}{dt}=\beta\aSI\left(1+z\frac{\aSI}{\aS}\right)-(2\gamma+\mu)\aII + K\mu\frac{\aI^2}{N^2},\nonumber\\
\fl\qquad\frac{d\aIR}{dt}=z\beta\frac{\aSI\aSR}{\aS}+\gamma\left(2\aII-\aIR\right)-\delta\aIR -\mu\aIR +2K\mu\frac{\aI\aR}{N^2},\nonumber\\
\fl\qquad\frac{d\aRR}{dt}=\gamma\aIR-2\delta\aRR -\mu\aRR +K\mu\frac{\aR^2}{N^2}.
\end{eqnarray}
Here every term is of order $N$ and lower order terms (including Gaussian stochastic corrections) have been neglected.\par

This system is similar to that studied in \cite{Shaw2008}, though there are several differences. Firstly, the random rewiring process has not been considered previously, and we also take a slightly different prescription for the smart rewiring process. More importantly, in most work using the PA, it is implicitly assumed that the average degree of susceptible nodes is simply $K/N$, not $\llbracket k_S\rrbracket$ as we use here.

With nine dimensions and seven parameters, the bifurcation structure of our deterministic model is likely to be complex. Indeed, analysis of a closely related model \cite{Shaw2008} reveals bifurcations of the transcritical, saddle-node and Hopf types. We are most interested in the presence of a Hopf bifurcation, whereby a stable fixed point becomes unstable to create a limit cycle. We can be sure that our model possesses such a transition, since in the limit of large $\mu$ it reduces to the usual well-mixed SIRS model, which does indeed have a Hopf bifurcation. For the remainder of the paper will restrict our attention to regions of parameter-space in which the deterministic model has a globally attractive endemic (meaning that $\aI\neq0$) fixed point $\bm{x}^\ast$. We aim to provide a description of the effects of demographic noise in this regime. 

\subsection{Linear noise approximation}
We next obtain a first-order approximation for the effects of noise following the standard procedure of linearising the Langevin equation (\ref{langevin}) around the endemic fixed point $\bm{x}^\ast$. In population-based models this procedure is equivalent to the van Kampen system size expansion \cite{vanKampen1992}. The central limit theorem suggests we should expect fluctuations of order $1/\sqrt{N}$, motiving a change of variables
\begin{equation}
\bm{\xi}=\sqrt{N}(\bm{x}-\bm{x}^\ast)\,.
\end{equation}
We apply this change to Eq. (\ref{langevin}) and retain only leading-order terms in $1/\sqrt{N}$ to obtain
\begin{equation}\label{eq:linlan}
\frac{d}{dt}\bm{\xi}={\mathbb A}\bm{\xi}+\bm{\zeta}(t)\,,
\label{lna}
\end{equation}
where ${\mathbb A}$ is the Jacobian matrix of the deterministic part of (\ref{langevin}), that is,
\begin{equation}
A_{ij}=\frac{\partial}{\partial x_j}\sum_k \mathbb{T}_{ik}r_{k}\bigg|_{\bm{x}=\bm{x}^\ast}\,,
\end{equation}
and $\bm{\zeta}$ is a vector of Gaussian noise variables whose correlation matrix ${\mathbb B}$ has entries 
\begin{equation}
B_{ij}=\sum_k\mathbb{T}_{ik}r_k\mathbb{T}_{kj}\bigg|_{\bm{x}=\bm{x}^\ast}\,,
\end{equation}
where $<\zeta_i(t)\zeta_j(t')>=B_{ij}\delta(t-t')$. If the fixed point is stable and the Jacobian $\mathbb{A}$ has complex eigenvalues, then the trajectories described by Eq. (\ref{lna}) may exhibit stochastic oscillations. To characterise this behaviour we take a Fourier transform, writing
\begin{equation}
\bm{\hat{\xi}}(\omega)=\frac{1}{\sqrt{2\pi}}\int \bm{\xi}(t)e^{- i \omega t}\,dt\,.
\end{equation}
In frequency space the object of interest is the power spectral density matrix, defined by $\mathbb{P}(\omega)=\int \big\langle \hat{\bm{\xi}}(\omega')\hat{\bm{\xi}}(\omega)^\dag\big\rangle\,d\omega'$, where $\big\langle \cdots\big\rangle$ denotes the ensemble average. For linear noise processes we have the explicit formula 
\begin{equation}
{\mathbb P}(\omega)= ({\mathbb A}-i\omega {\mathbb I})^{-1}{\mathbb B}({\mathbb A}^T+i\omega{\mathbb I})^{-1},
\label{specmatrix}
\end{equation}
which can be derived directly from Eq. (\ref{eq:linlan}) and using the properties of the noise $\bm{\zeta}$. The quantity ${\mathbb I}$ is here the $9\times 9$ identity matrix.
The diagonal entries of ${\mathbb P}(\omega)$ contain the power spectrum of oscillations for each of the entries of $\bm{\xi}$. For example, fluctuations in the number of susceptible agents are described by $P_{1,1}(\omega)$. The off-diagonal entries of ${\mathbb P}(\omega)$ give information on the correlations between state variables. For example, oscillations in the density of susceptible agents are typically out of phase with those of infective agents, leading to complex values of $P_{1,2}(\omega)$ with argument close to $\pi/2$ \cite{Rozhnova2012}. \par

\section{Test against simulations}

\begin{figure}
\includegraphics[width=\textwidth]{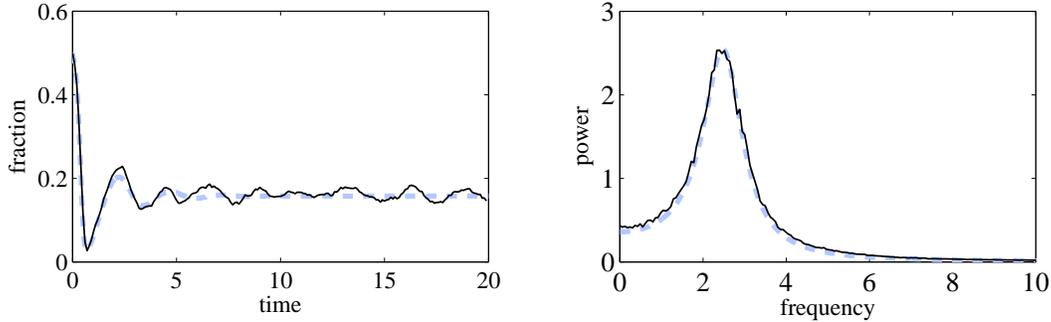}
\caption{Left: time series of the fraction of susceptible agents coming from solution of the PBP deterministic equations (\ref{deteqns}) (thick dashed blue line) and a simulation of the adaptive network (thin solid black line). Right: Power spectrum of oscillations in the fraction of susceptible agents as computed for the PBP, see Eq. (\ref{specmatrix}) (thick dashed blue line), and measured from simulations of the adaptive network (thin solid black line). The parameters used were $\beta=6$, $\delta=3$, $\gamma=0.5$, $\mu=5$ $w=30$, $N=10000$, $K=100000$.}
\label{sfig}
\end{figure}

As with all approximation schemes of this type, the motivation for creating the PBP was to provide theoretical predictions that are a reasonable match to the behaviour of the adaptive-network model in simulations. In Fig.~\ref{sfig}, we show an example of the comparison between theoretical results of the PBP and data gathered from a corresponding simulation of the adaptive-network model (performed using Gillespie's algorithm \cite{Gillespie1977}). As shown in Fig.~\ref{sfig}, the agreement is generally good, though we should point out that there is a particular area of parameter space in which there are significant discrepancies between the models. When both the mean degree $K/N$ and the rate of random rewiring $\mu$ are small, the typical states of the adaptive-network model are highly heterogeneous across the network. This heterogeneity makes a description of the network only in terms of pairs too weak to capture the details of the model, and the agreement with the PBP breaks down, particularly in the transient part of the dynamic. 

\begin{figure}
\includegraphics[width=\textwidth]{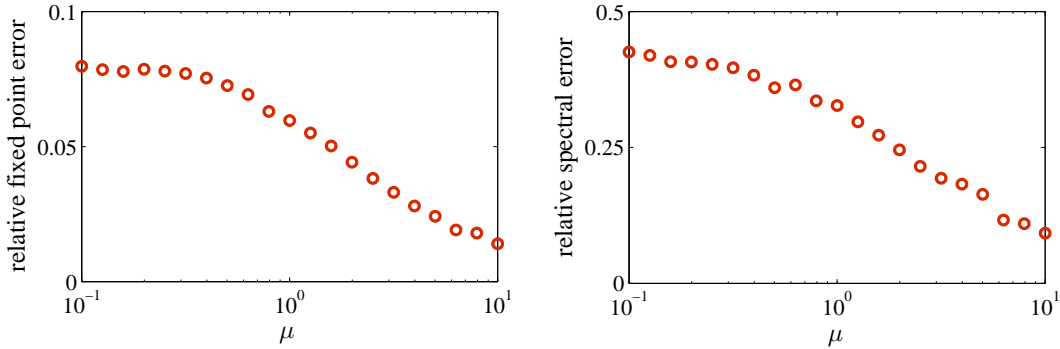}
\caption{Discrepancy between the PBP and adaptive network simulations, as a function of random rewiring rate $\mu$. The left hand plot displays the relative error between the equilibrium fraction of susceptible individuals in the PBP and adaptive network simulations. In the right hand plot a measure for the relative error in the power spectrum of oscillations of susceptible individuals is computed by integrating (from $\omega=0$ to $\omega=10$) the difference between the theoretical result of the PBP and the spectrum measured from the network simulations, and dividing by the total power of the simulation result in that frequency domain. The other model parameters were (arbitrarily) chosen as $\beta=6$, $\delta=3$, $\gamma=0.4$, $w=4$, $N=1000$, $K=10000$.}
\label{errfig}
\end{figure}

On a mathematical level, the root cause of the difficulty in approximating adaptive-network models with low-dimensional systems is the development of long-range correlations within the network. Heuristically, it is easy to imagine that the random rewiring process we have introduced will act to curtail the development of long-range correlations and move the behaviour of the model closer to that of a well-mixed population. Whilst the derivation of the PBP did not involve anything so exact as a formal expansion in $1/\mu$, we expect that the PBP and the adaptive network will agree in the limit $\mu\to\infty$. We have tested this prediction by varying $\mu$ over two orders of magnitude and measuring the agreement between network simulations and the PBP theory; the results are shown in Fig. \ref{errfig}.

\section{Oscillations of the network structure}
It is well-known in the literature on epidemic modelling that demographic stochasticity can excite periodic fluctuations in the prevalence of the disease, leading to cyclical outbreaks (see, for example, \cite{Rozhnova2011,Alonso2007,Kuske2007,Rozhnova2009}). It is natural then to ask if the structure of the network of infectious contact also undergoes stochastic oscillations in response to the stochasticity of the disease dynamic. 

\begin{figure}
\includegraphics[width=\textwidth]{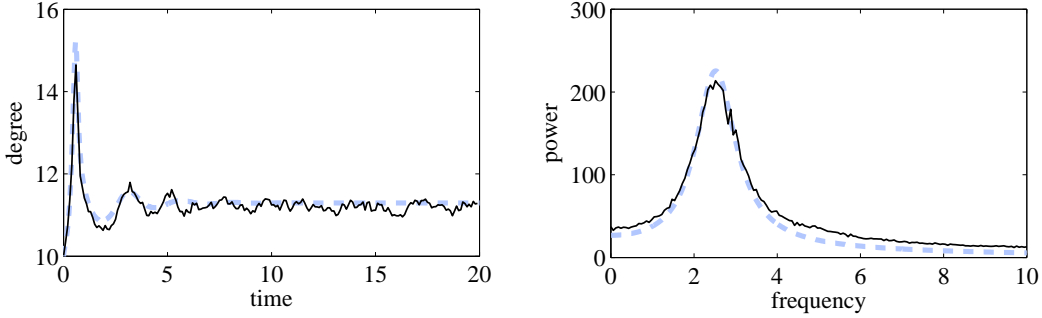}
\caption{Left: time series of the average degree of susceptible agents coming from solution of the PBP deterministic equations (\ref{deteqns}) (thick dashed blue line) and a simulation of the adaptive network (thin solid black line). Right: Power spectrum of oscillations in the average degree of susceptible agents as computed for the PBP (thick dashed blue line), and measured from simulations of the adaptive network (thin solid black line). The parameters used were the same as those in Fig~\ref{sfig}, namely, $\beta=6$, $\delta=3$, $\gamma=0.5$, $\mu=5$ $w=30$, $N=10000$, $K=100000$.}
\label{skfig}
\end{figure}

This information is contained in the oscillations of the mean degree of susceptible nodes, $[k_S]$, in the adaptive-network model, which have amplitude of the order of $1/\sqrt{N}$. The power spectrum for $\llbracket k_S\rrbracket$ (the PBP model's equivalent of $[k_S]$), can be derived from that of the other variables of the PBP. Recalling that
\begin{equation}
k_S= \frac{2\aSS+\aSI+\aSR}{\aS}=\frac{2x_4+x_5+x_6}{x_1}\,,
\end{equation}
we linearise around the deterministic fixed point value $k_S^\ast$ to obtain
\begin{equation}
\kappa_S=\sqrt{N}\Big(k_S-k_S^\ast\Big)\approx \frac{2\xi_4+\xi_5+\xi_6-k_S^\ast\xi_1}{x_1^\ast}\,.
\end{equation}
Carrying out the Fourier transform is straightforward, and gives
\begin{equation}
\hat{\kappa}_S(\omega)=\frac{2\hat{\xi_4}(\omega)+\hat{\xi_5}(\omega)+\hat{\xi_6}(\omega)-k_S^\ast\hat{\xi_1}(\omega)}{x_1^\ast}\,,
\end{equation}
and thus the power spectrum of oscillations in $k_S$ is computed as
\begin{eqnarray}
\fl\big\langle|\hat{\kappa}_S(\omega)|^2\big\rangle=\left(\frac{1}{x_1^\ast}\right)^{2}\Bigg\{(k_S^\ast)^2&P_{1,1}(\omega)+ 4P_{4,4}(\omega) + P_{5,5}(\omega) + P_{6,6}(\omega)\nonumber \\&+ 4\re\big[ P_{4,5}(\omega)\big]  +4\re\big[ P_{4,6}(\omega)\big]  +2\re\big[ P_{5,6}(\omega)\big] \\&-k_S^\ast \Big(4\re\big[ P_{1,4}(\omega)\big] +2\re\big[ P_{1,5}(\omega)\big] +2\re\big[ P_{1,6}(\omega)\big] \Big)\Bigg\}\,.\nonumber
\end{eqnarray}
In Fig.~\ref{skfig} we show a comparison between the deterministic dynamics and stochastic oscillations in $\llbracket k_S\rrbracket$ in the PBP to those of $[k_S]$ in simulations of the adaptive-network model. The parameter values are the same as those used in Fig \ref{sfig}. Whilst the agreement in both plots is not quite as close as in the comparison of $\aS$ to $[S]$, the PBP still clearly provides a strong indication of the oscillations in network structure. 

\begin{figure}
\includegraphics[width=\textwidth, trim=0 20 0 0]{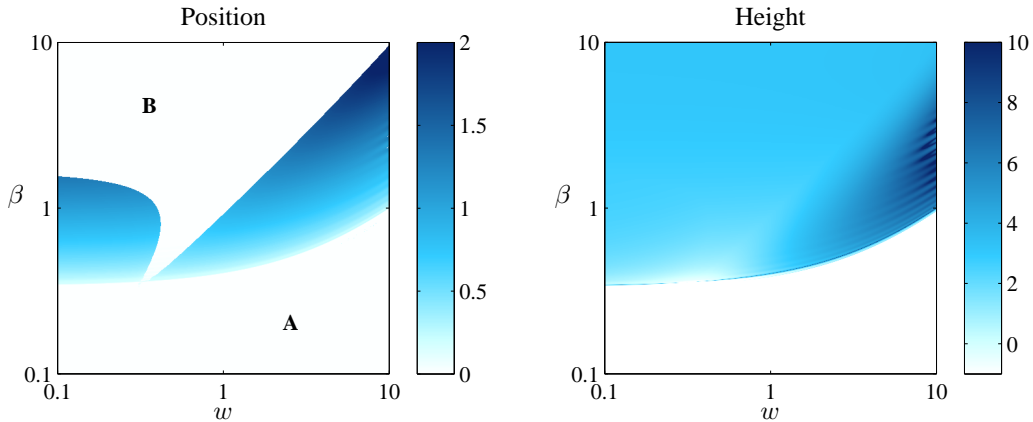}
\caption{Colourmaps showing the position and height (on a logarithmic scale) of the peak in the power spectrum of oscillations in $\llbracket k_S\rrbracket$, as a function of infection rate $\beta$ and smart rewiring rate $w$, both varying on a logarithmic scale from 0.1 to 10. The other parameters are $\delta=3$, $\gamma=0.2$, $\mu=2$, $K/N=10$. In region A the trivial (i.e. disease extinction) steady state is attractive and no fluctuations are present. Region B denotes the area of parameter space in which the stochastic oscillations are present, but the spectrum is dominated by the zero mode; time series from this region are difficult to distinguish from pure white noise.}
\label{posheifig}
\end{figure}
It is interesting to explore how the expression of stochastic oscillations in network structure varies with parameters. The colourmaps in Fig.~\ref{posheifig} show the position and height of the peak in the spectral density, as the infection rate $\beta$ and smart rewiring rate $w$ are varied over two orders of magnitude. Several interesting features are visible in the figures. At low values of $\beta$ there is a region (marked A in the figure) in which the disease typically dies out and thus no fluctuations are present. Two behaviours are possible in the active phase. In region B, stochastic oscillations are present, but the spectrum is dominated by a peak at the origin which represents the excitation of low frequency modes (although other, smaller, peaks elsewhere in the spectrum are still possible). Outside of regions A and B the power spectrum has its global maximum at a non-zero frequency; here the oscillations generally become faster as $\beta$ is increased and more powerful as $w$ is increased. In principle, the transition in the power spectrum from being dominated by a peak at the origin to a peak at non-zero frequency $\omega$ could happen in one of two ways: either the peak at zero could migrate, or a secondary peak could grow elsewhere in the spectrum and eventually overtake the one at the origin. Since the boundary of region B is sharp, we can conclude that in our model the second mechanism is at work.

\section{Conclusions}
To summarize, we have put forward an approximative modelling approach with which to study the effects of intrinsic fluctuations in interacting agent systems on adaptive networks. Our method follows a similar line to that of \cite{Rozhnova2009}; the key step consists of deriving an `effective' stochastic process for the average numbers of single and pairs of agents in each possible state. This technique was previously seen to be viable for static regular networks, we here show that it can be successful for the case of co-evolving adaptive networks as well. We define the dynamical rules of our pair-based proxy model to follow those of the original network model as closely as possible, based on the typical changes that occur in the network model during infection, recovery, birth or rewiring events. 

Our approach relies on an approximation that neglects long-range correlations within the network, equivalent to the usual PA which is often employed when writing down deterministic effective descriptions. As such, the behaviour of the PBP may only be considered an accurate representative of the full adaptive-network model in situations where typical states of the network are not too spatially heterogeneous. The agreement is aided in our model by the presence of a low-level random rewiring process which acts to `stir' the network. As this rate is increased we observe a corresponding improvement in the accuracy of the predictions made by the PBP. As well as improving the predictive power of our methods, we would argue that the presence of background of random rewiring is not unrealistic as we expect that most naturally occurring networks are subject to extrinsic factors (e.g. movement of agents) which randomise their structure. 

Comparisons with network simulations demonstrate that the PBP is able to describe the behaviour of the underlying adaptive-network model on the aggregate deterministic level (fixed points are reproduced to a good accuracy), and crucially also on the level of Gaussian fluctuations for large, but finite networks. To our knowledge no previous (semi-) analytical work exists which would predict the power spectra of quasi-cycles in adaptive-network models to the accuracy shown in Fig. \ref{sfig}. While our work focuses on the specific application to models of epidemic spread we expect that they can be transferred to adaptive-network models of other processes as well, in particular for example models in game theory \cite{Szabo, demirel,zschaler}, opinion dynamics \cite{vazquez} or ecology \cite{Proulx}.

Almost as a by-product we also obtain access to oscillatory behaviour of the network itself. In the adaptive-network model the dynamics {\em on} and {\em of} the network co-evolve, and as such one would expect the network structure itself to oscillate in regimes in which the disease dynamics shows quasi-cycles. This is indeed confirmed by simulations, in which we observe noise-driven oscillations with amplitude proportional to $1/\sqrt{N}$. Moreover, our analysis using the PBP allows us to obtain accurate results for the spectral properties of these oscillations in network structure. Since our approach is analytical, we are able to systematically explore parameter space, for example to determine the relationship between the character of the oscillations and the rates of infection and smart rewiring. We expect that other adaptive-network models will also exhibit fluctuations in their structure which can be captured within our approach. 

\section*{Acknowledgements}
TR acknowledges funding by the EPSRC under grant number EP/H02171X/1. TG is supported by RCUK (reference EP/E500048/1), and by the EPSRC (references EP/I005765/1 and EP/I019200/1). The authors would like to thank Thilo Gross and Ganna Rozhnova for helpful discussions.
\section*{References}
\bibliographystyle{iopart-num}
\bibliography{Adaptive_refs}

\end{document}